%% file: main.tex
\documentclass[sigconf]{acmart}

\AtBeginDocument{%
  }

\setcopyright{acmcopyright}
\copyrightyear{2024}
\acmYear{2024}
\acmDOI{XXXXXXX.XXXXXXX}

\acmConference[ICPC 2024]{46th International Conference on Program Comprehension}{April 2024}{Lisbon, Portugal}

\acmBooktitle{46th International Conference on Program Comprehension (ICPC 2024),
 April, 2024, Lisbon, Portugal}
\acmPrice{15.00}
\acmISBN{978-1-4503-XXXX-X/18/06}

\usepackage{graphicx}
\usepackage{verbatim}

\usepackage{colortbl}
\newcommand{\colorblue}{\cellcolor[rgb]{0.757,0.867,1}}
\newcommand{\coloryellow}{\cellcolor[rgb]{1,0.925,0.792}}

\usepackage{arydshln}

\usepackage{multirow}
\usepackage{enumitem}

\usepackage{fancyhdr}
\usepackage[framemethod=TikZ]{mdframed}
\usepackage{tcolorbox}
\makeatletter
\newcommand{\mybox}[1]{%
  \setbox0=\hbox{#1}%
  \setlength{\@tempdima}{\dimexpr\wd0+13pt}%
  \begin{tcolorbox}[boxrule=0.5pt, colback=gray!10, arc=4pt,
      left=6pt,right=6pt,top=6pt,bottom=6pt,boxsep=0pt]
    #1
  \end{tcolorbox}
}

\newcommand{\find}[1]{
\begin{tcolorbox}[leftrule=1mm,toprule=0mm,bottomrule=0mm,left=1pt,right=2pt,top=2pt,bottom=2pt]
\em #1
\end{tcolorbox}
}

\usepackage{subcaption}

\begin{document}

\title{Knowledge-Aware Code Generation with Large Language Models}

\author{Tao Huang}
\affiliation{%
 \institution{School of Information Science and Engineering, Shandong Normal University}
 \state{Jinan}
 \country{China}
 }
\email{2022317095@stu.sdnu.edu.cn}

\author{Zhihong Sun}
\affiliation{%
 \institution{School of Information Science and Engineering, Shandong Normal University}
 \state{Jinan}
 \country{China}
 }
\email{2022021002@stu.sdnu.edu.cn}

\author{Zhi Jin}
\affiliation{%
 \institution{Key Lab of HCST (PKU), MOE; SCS, Peking University}
 \state{Beijing}
 \country{China}
 }
\email{zhijin@pku.edu.cn}

\author{Ge Li}
\affiliation{%
 \institution{Key Lab of HCST (PKU), MOE; SCS, Peking University}
 \state{Beijing}
 \country{China}
 }
\email{lige@pku.edu.cn}

\author{Chen Lyu}
\authornote{This work was done when Chen Lyu was a visiting scholar at Peking University.}
\affiliation{%
 \institution{School of Information Science and Engineering, Shandong Normal University}
 \state{Jinan}
 \country{China}
 }
\email{lvchen@sdnu.edu.cn}

\renewcommand{\shortauthors}{Huang, et al.}

\input{sec/0.abstract}





\maketitle

\input{sec/1.introduction}
\input{sec/2.relatedwork}
\input{sec/3.studydesign}
\input{sec/4.approach}
\input{sec/5.environmentsetup}
\input{sec/6.resultandanalysis}
\input{sec/7.threats_to_validity}

\input{sec/7.disscussion}
\input{sec/8.conclution}

\bibliographystyle{ACM-Reference-Format}

\end{document}

%% file: sec/0.abstract.tex
\begin{abstract}
Large Language Models (LLMs) perform well on basic programming problems. However, they encounter challenges when dealing with complex tasks involving the use of diverse algorithmic and data structure skills, particularly programming competition-level problems. Notably, ChatGPT exhibits proficient performance on problems it has encountered during its pre-training phase, but this performance deteriorates when faced with novel problems. Consequently, enhancing the ability of LLMs to address unfamiliar problems has emerged as a pivotal research focus. The problem-solving process of LLMs mirrors human programmers' approach to a certain extent. When confronted with new programming tasks, human programmers engage in task planning and code writing with the previously acquired knowledge about algorithms and data structures. Despite having learned such knowledge, LLMs struggle to effectively apply it when faced with specific new problems. To address this issue, we constructed a novel dataset, CodeF, which contains a portion of programming problems that ChatGPT has not previously encountered. Furthermore, we developed a Knowledge Library tailored for Python programming contest problems and introduced the concept of \textbf{K}nowledge-Aw\textbf{are Code} Gene\textbf{r}ation (KareCoder). KareCoder bolsters the models' understanding and problem-solving capabilities by integrating prompt and knowledge from the library into the LLMs' code generation reasoning process, especially on Pass@1 metrics. Upon testing on the CodeF and APPS datasets, KareCoder demonstrated outstanding performance in handling novel problems previously unencountered by LLMs. In contrast with the code directly generated by ChatGPT, KareCoder achieved a relative improvement of 23.3\% on the Pass@1 metric on the  CodeF post2021-9 dataset. Additionally, it performs well compared to other methods when dealing with problems that LLMs have previously encountered. Our dataset and experiment data are open-sourced and can be accessed at \url{https://github.com/CodeGeneration3/KareCoder}.
\end{abstract}

%% file: sec/1.introduction.tex
\section{Introduction}\label{sec:1}
Code generation tasks aim to automatically generate executable programs based on natural language descriptions. In recent years, code generation tasks have garnered substantial attention and undergone extensive development in both the academic and industrial realms. Some prominent applications encompass \cite{r1poesia2021synchromesh,r2shen2022incorporating,r3chen2022codet,r4dong2023codep,r36lyu2021embedding}. Specifically, the recent proliferation of Large Language Models (LLMs) such as CodeGen \cite{r5nijkamp2022codegen}, CodeX \cite{r6chen2021evaluating} and ChatGPT \cite{r7OpenAI2022}, has not only induced a profound impact on the domain of code generation but also significantly fostered the progress of associated fields, including Natural Language Processing (NLP) and intelligent software engineering.

\begin{figure}[h]
  \centering
  \includegraphics[width=\linewidth]{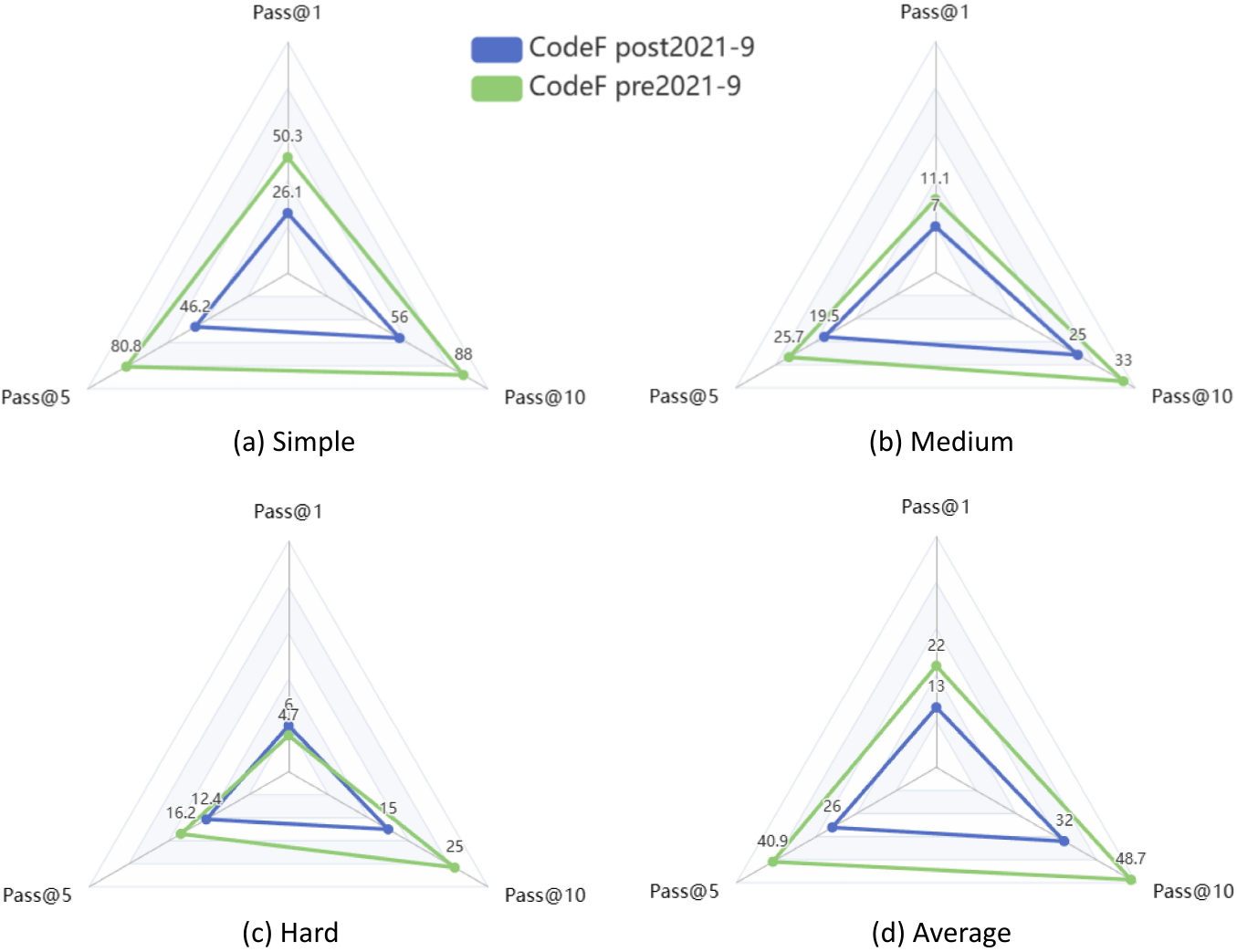}
  \caption{Experimental validation of split-difficulties on CodeF post2021-9 and CodeF pre2021-9 datasets.}
  \label{fig:radermap}
\end{figure}

Code generation can be conceptualized as a complex reasoning task. The objective of this task is to transfigure Natural Language (NL) descriptions into Programming Language (PL) forms. Typically, this process encompasses two phases: training and inference, requiring a copious number of natural language-code pairs (<Text, Code>) as a foundation. High-quality datasets play a pivotal role in the code generation task. Several related studies \cite{r8lee2020biobert,r9gururangan2020don} have demonstrated that with the enlargement of the model size, augmentation in the quantity of data and escalation in computational prowess, the performance of the model accordingly experiences enhancement. Nevertheless, empirical evidence suggests that utilizing the model's training set as a test set for inference may lead to over-optimized results. Given that the training data for LLMs are not publicly accessible, datasets frequently employed in code generation tasks, such as MBPP \cite{r10austin2021program}, Humaneval \cite{r6chen2021evaluating} and APPS \cite{r12hendrycks2021measuring}, may be encompassed in the training set of LLMs. Consequently, should we use these datasets as test data, whether the genuine test outcomes would be compromised due to this potential data overlap constitutes an issue warranting our thorough exploration.

Based on the above thinking, we construct CodeF to investigate the potential overlap of test data included in the training data of ChatGPT. This research aims to decipher whether this overlap could result in an inflation of actual results. We employ the ``gpt-3.5-turbo-0613'' model of ChatGPT and use the cutoff time (September 2021) of ChatGPT's training data \cite{r13OpenAI2023} as disclosed by OpenAI. This date serves as the limit to divide our dataset CodeF\footnote{CodeF is a novel dataset we have independently designed to meet task requirements, resolving the issue of potential data obsolescence that may exist in extant datasets (specifically, the data may be incorporated in the training data of LLMs like ChatGPT). Detailed information of the dataset can be found in the ``Dataset Collection'' subsection in Section \ref{sec:3.1}. }. We used a direct generation approach with ChatGPT to produce codes for both parts of the dataset and evaluated the results using Pass@1, Pass@5 and Pass@10. The results are shown in Figure \ref{fig:radermap}. As Figure \ref{fig:radermap} demonstrates, the experimental results on CodeF pre2021-9 outperform those on CodeF post2021-9, especially regarding problems of easy difficulty, where the relative difference of Pass@1 between the two reaches an astonishing 92.7\%. The somewhat lower Pass@1 score for hard difficulty problems in CodeF pre2021-9 might be due to the excessive complexity of these problems, ChatGPT may not have successfully captured the features of these problems.

Drawing on the aforementioned results, LLMs have indeed acquired knowledge of problems they encountered during the training phase. Nonetheless, they appear to lack a strategy for addressing unencountered problems. This results in the disparity in outcomes between the two parts of data depicted in Figure \ref{fig:radermap}. The situation is akin to a student sitting for an examination: they may effortlessly solve problems they have previously reviewed, yet struggle with new and unreviewed problems. With the tutors provide guidance, the student has a significantly enhanced likelihood of successfully completing the problems. We know that algorithms and structures serve as important features of code, so can code generation be enhanced by incorporating algorithms and structures? Hence, we consider infusing some knowledge into the problems in the assistance of in-context learning. This approach can serve to augment ChatGPT's relevant knowledge, thereby ameliorating its proficiency in solving previously unencountered problems. 

In previous research, Jiang et al. \cite{r14jiang2023self} put forth a code generation method termed ``Self-planning''. This technique harnesses the innate capability of LLMs to strategize the programming problem, thereby guiding the code generation process. Concurrently, Li et al. \cite{r15li2023think} proposed a method called ``Brainstorming Boost'', aimed at stimulating LLMs for deeper introspection. Dong et al. \cite{r16dong2023self} employed a multi-role interaction model to facilitate cooperation and interaction amongst LLMs during code generation tasks. Drawing inspiration from these methods, we opted to enhance the input information fed into LLMs to improve their inferential capabilities when addressing previously unencountered problems. This in turn improves their generalization capacities for handling new programming problems. Consequently, we are introducing a new approach named \textbf{K}nowledge-Aw\textbf{are Code} Gene\textbf{r}ation (KareCoder).

Specifically, we first organized and built a Knowledge Library involving algorithms and data structures information based on the tags of algorithms and data structures that may be used in complex programming problems. Subsequently, KareCoder exploits the planning capabilities of LLMs in conjunction with the external Knowledge Library to direct code generation. The operational process of KareCoder unfolds over two distinct stages:

\begin{itemize}
\item {\textbf{Prompt Engineering Stage:}} In this stage, the LLMs learn the one-shot example prompt that we furnish and subsequently generate knowledge-aware prompt pertinent to the problem based on the algorithms and data structures information in the Knowledge Library. 
\item {\textbf{Coding Stage:}} In this stage, we guide the LLMs to incorporate the problem and the prompt generated in the preceding step. Then, under the guidance of the prompt, the model systematically generates code which addresses the corresponding programming problems. 
\end{itemize}

In summary, the main contributions of this paper are as follows:

\begin{enumerate}[label=(\arabic*)]
  \item We constructed a novel code generation dataset, CodeF, consisting of problems manually crawled, cleaned and inspected to mitigate the issue of data overlap with the training data of LLMs. For each programming problem, we extracted pertinent information relating to the algorithms and data structures tags, difficulty and date (time of release). 
  \item We assembled a Knowledge Library for Python programming problems, detailing the algorithms and data structures potentially employed in resolving programming problems. We propose a approach grounded in LLMs, named KareCoder, which amalgamates algorithms and data structures knowledge into the code generation process. 
  \item We conducted a comprehensive validation of the efficacy of KareCoder on both CodeF and APPS benchmarks. KareCoder surpasses other competitors on CodeF (e.g., ChatGPT \cite{r7OpenAI2022}, Self-planning \cite{r14jiang2023self}, SCOT \cite{r31li2023enabling}, SCOT\&KareCoder), while also demonstrating commendable performance on APPS in comparison to other methods outside of ChatGPT. 
\end{enumerate}

%% file: sec/2.relatedwork.tex
\begin{figure*}
  \centering
  \includegraphics[width=\linewidth]{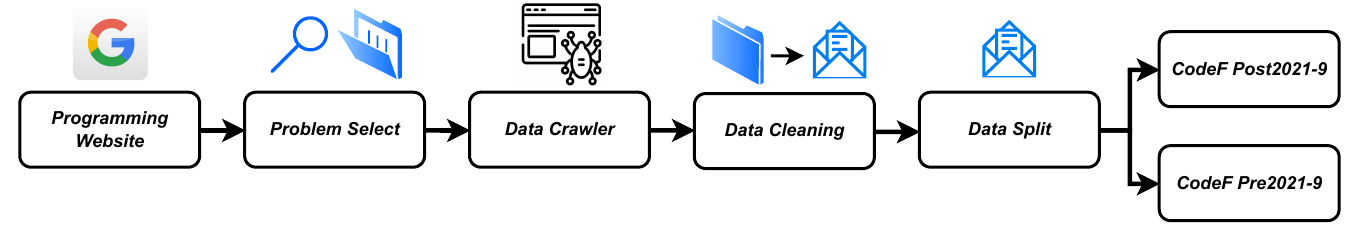}
  \caption{Schematic diagram of CodeF acquisition and processing. The processing part includes data cleaning and splitting.}
  \label{3.1datasetprocess}
\end{figure*}
\section{Related Work}\label{sec:2}

\subsection{Code Generation}\label{sec:2.1}

Model size divides code generation tasks into those using small to medium models and LLMs. Ling et al. \cite{r17ling2016latent} first introduced a method to translate natural language into code fragments utilizing a sequence-to-sequence model. Sun et al. \cite{r18sun2020treegen} adopted the Transformer architecture to resolve the issue of long-term dependencies amongst code elements, proposing TreeGen to enhance the model and incorporate information regarding the code structure. Wang et al. \cite{r21wang2021codet5} proposed CodeT5, which integrates the code's features during the pre-training phase, emphasizing the model's reasoning about tokens with practical significance during the generation phase. Dong et al. \cite{r4dong2023codep} devised a PDA-based approach to guarantee the syntactic correctness of code generation.

As the volume of training data for language models proliferates, more LLMs are being utilized for code generation. The emergence of CodeBERT \cite{r20feng2020codebert} has marked a new epoch in the use of pre-trained LLMs. Pre-trained LLMs such as CodeGen \cite{r5nijkamp2022codegen}, InCoder \cite{r22fried2022incoder}, CodeX \cite{r6chen2021evaluating}, AlphaCode \cite{r23li2022competition}, Code Llama \cite{r33roziere2023code} and StarCoder \cite{r32li2023starcoder} have yielded significant enhancements in code generation performance. Beyond these models tailored for code-related tasks, ChatGPT \cite{r7OpenAI2022}, a model designed for problem-and-answer applications, has also demonstrated remarkable code generation capabilities.

\subsection{Code Generation Dataset}\label{sec:2.2}
MBPP \cite{r10austin2021program} and HumanEval \cite{r6chen2021evaluating} are datasets commonly utilized in code generation related research \cite{r3chen2022codet,r14jiang2023self,r16dong2023self}. They comprise hand-written programming problems, which are intrinsically simple. They include only rudimentary task descriptions and provide relatively short solutions, thereby not meeting the complexity required for our daily tasks. Conversely, datasets like APPS \cite{r12hendrycks2021measuring}, derived from several open-source programming competition websites, offer considerably more challenging and lengthier problems, thus serving as a more objective measure. CodeContest \cite{r23li2022competition}, designed to fine-tune and evaluate the AlphaCode model, has observed a significant enhancement in the quality of test cases compared to previous datasets. Despite the satisfactory performance of these datasets, there exists a need for a new dataset that will not overlap with the training data of ChatGPT. The cutoff time for the training data of ChatGPT, which OpenAI has made public \cite{r13OpenAI2023}, is September 2021. In order to circumvent a substantial count of false positives in the training data when working with LLMs, there is an immediate requirement for a new dataset encompassing data from September 2021 onwards. Furthermore, the dataset also needs to include the algorithms and data structures tags recommended for the problems to enable more effective integration of algorithms and data structures information. 

\subsection{Prompt Techniques}\label{sec:2.3}
\sloppy{As LLMs grow in size and parameters, fine-tuning them needs more resource and time. Numerous recent studies\cite{r25liu2023pre,r26qiao2022reasoning,r28wang2023plan} have explored enhancing the performance of LLMs by integrating prompts. For instance, Zhang et al. \cite{r27zhang2022automatic} introduced an automated COT prompting method, AutoCoT, which samples diverse problems and generates inference chains to construct examples. Concurrently, Zhou et al. \cite{r30zhou2022least} proposed a least-to-most prompting strategy, where complex problems are segmented into a series of sub-problems that are then sequentially addressed. Similarly, Liu et al. \cite{r29liu2023improving} delved into the feasibility of guiding ChatGPT in code generation tasks through manually crafted prompts. By incorporating some external information, we could offer prompts to LLMs, thereby enabling them to generate superior prompts for task planning. }

%% file: sec/3.studydesign.tex
\section{Support Data Design}\label{sec:3}

In this section, we delineate the dataset CodeF and Knowledge Library that we assembled. Specifically, we expound on the reasons behind constructing a new dataset for a programming competition problem, the sources of CodeF and the three distinct types of Knowledge Libraries that we formulated. 

\subsection{Dataset Collection}\label{sec:3.1}
Models are conventionally divided into two crucial phases: training and testing. Our research aims to explore the intricate interplay between algorithms, data structures and the overlap between training and testing sets, and their influence on the code generation capability of large models. We undertook the creation of a specialized dataset for an algorithm-centered programming competition, which we named CodeF. This dataset is informed by the methodologies and insights gained from creating the TACO dataset \cite{r35li2023taco}. Taking September 2021 (the cutoff date for ChatGPT's training data \cite{r13OpenAI2023}) as the node, we divided CodeF into two parts: CodeF pre2021-9 whose problems ChatGPT might have encountered during its training, CodeF post2021-9 which consists of problems that ChatGPT would not have previously seen. 

In this study, we have constructed a new dataset of Python programming problems, CodeF, which incorporates 1523 problems posted from January 2020 to April 2023 on the CodeForces\footnote{\url{https://codeforces.com}} programming website. As depicted in Figure \ref{3.1datasetprocess}, the creation of CodeF primarily involves two phases: data acquisition and processing. 

During the data acquisition phase, we evaluated several programming contest platforms, including Codeforces, HackerRank and Geeksforgeeks. Factors such as legal restrictions, interface complexities of the platforms and our specific requirement for algorithms and data structures tag for the problems, as well as the release time information, led us to select CodeForces (a programming competition website) as our data source. In order to ensure dataset quality, we designed an HTML parser specifically tailored for the CodeForces site. We crawled the site for problem description, solution, input and output examples and a set of associated labels, which include difficulty, date and tag (indicating algorithms and data structures suitable for solving the problem), etc. We consolidated all the problems into a standardized format.

During the data acquisition process, we endeavoured to ensure the consistency of our data with that of the CodeForces website. However, since the code was sourced from user submissions, often embedded with comments and potential error codes, we implemented the following measures during the data processing phase:

\begin{itemize}
\item \textbf{Code De-commenting:} For a cleaner CodeF and to remove comments that are considered invalid information, we employed Abstract Syntax Tree (AST) parsing to remove comment nodes from code that was heavily annotated. 

\item  \textbf{Code De-duplication:} We collected codes from various users, leading to potential duplicates. Using the MinHash algorithm and Jaccard index, with a threshold of 0.85, we minimized these duplicates to unique files.

\item \textbf{Unit Test Validation:} Although we extracted code that was verified as correct by the website, we conducted unit tests on all codes to prevent errors during parsing. This step assured the accuracy of the test cases and codes. 
\end{itemize}

\input{table/datasetdifficulty}

\input{table/dataset-condition-new}

\begin{figure}[h]  
    \centering
    \begin{subfigure}[b]{0.45\textwidth}
        \includegraphics[width=\textwidth]{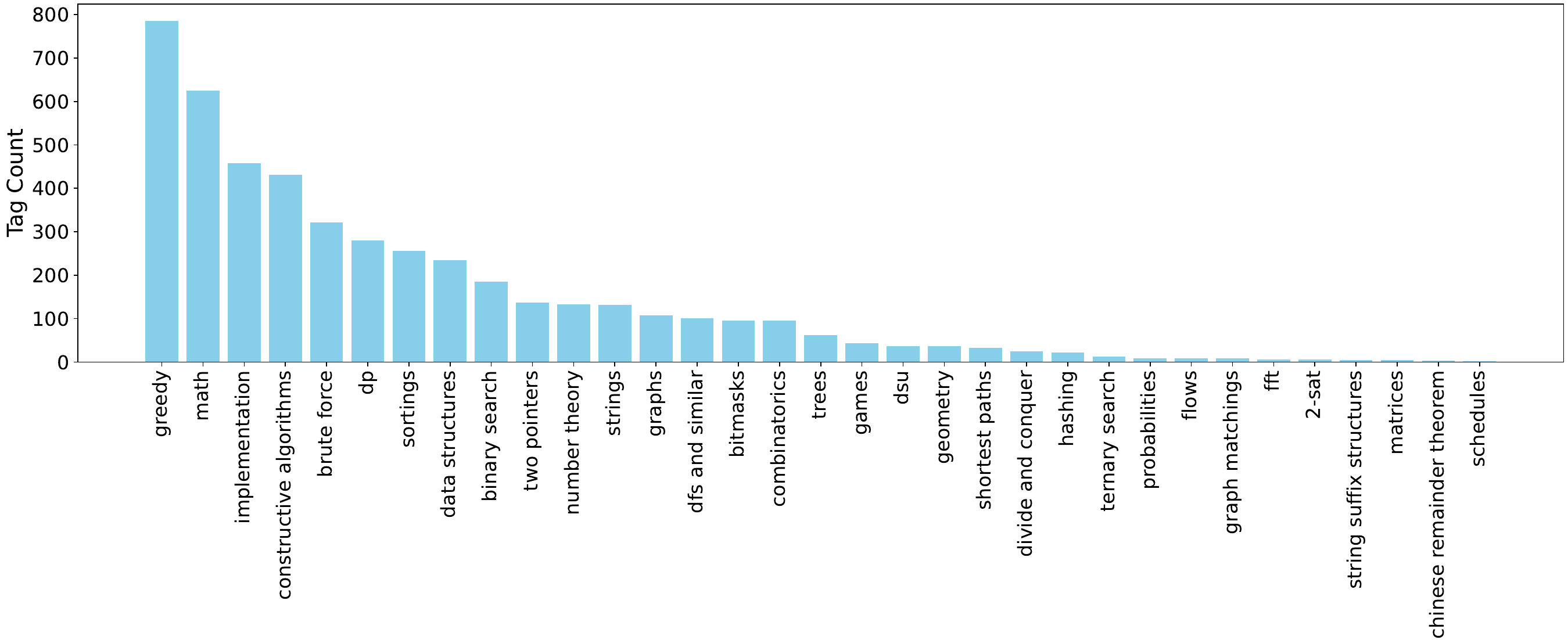}
        \caption{Distribution of CodeF}
        \label{fig:sub1}
    \end{subfigure}
    \hfill
    \begin{subfigure}[b]{0.45\textwidth}
        \includegraphics[width=\textwidth]{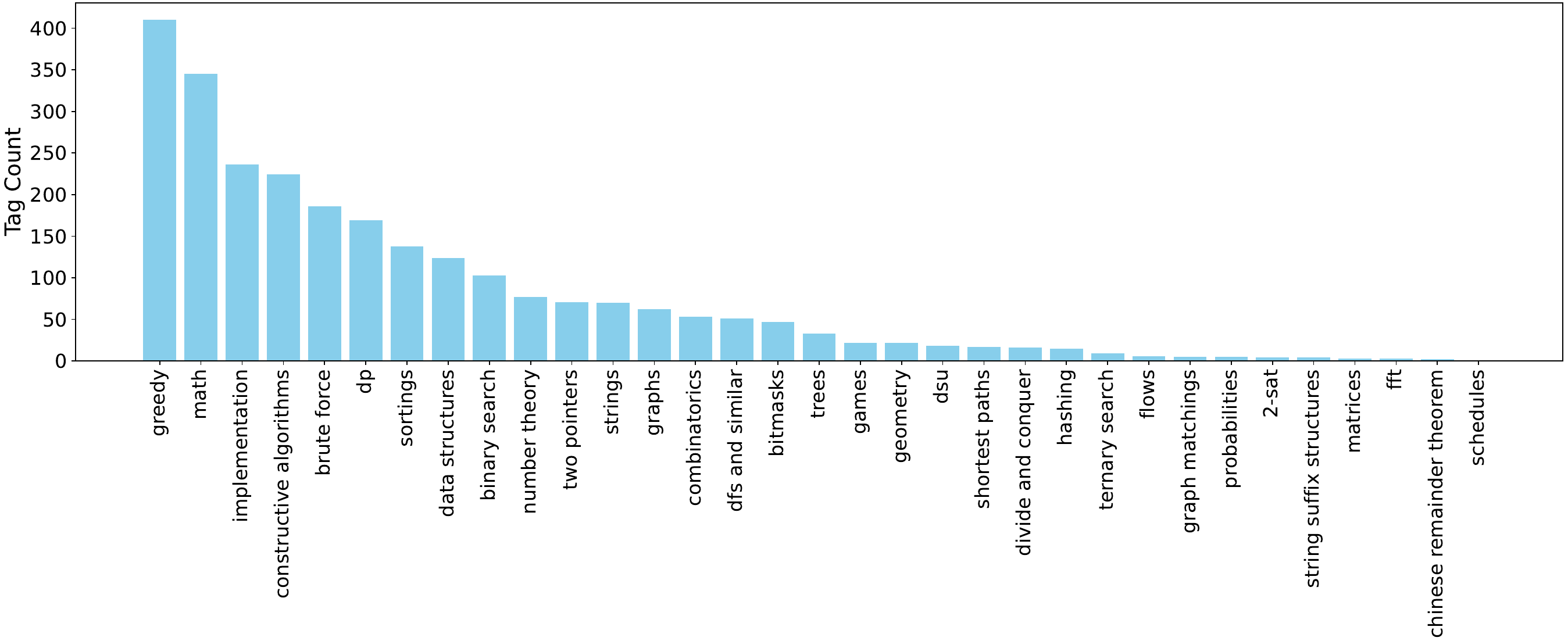}
        \caption{Distribution of CodeF pre2021-9}
        \label{fig:sub2}
    \end{subfigure}
    \hfill
    \begin{subfigure}[b]{0.45\textwidth}
        \includegraphics[width=\textwidth]{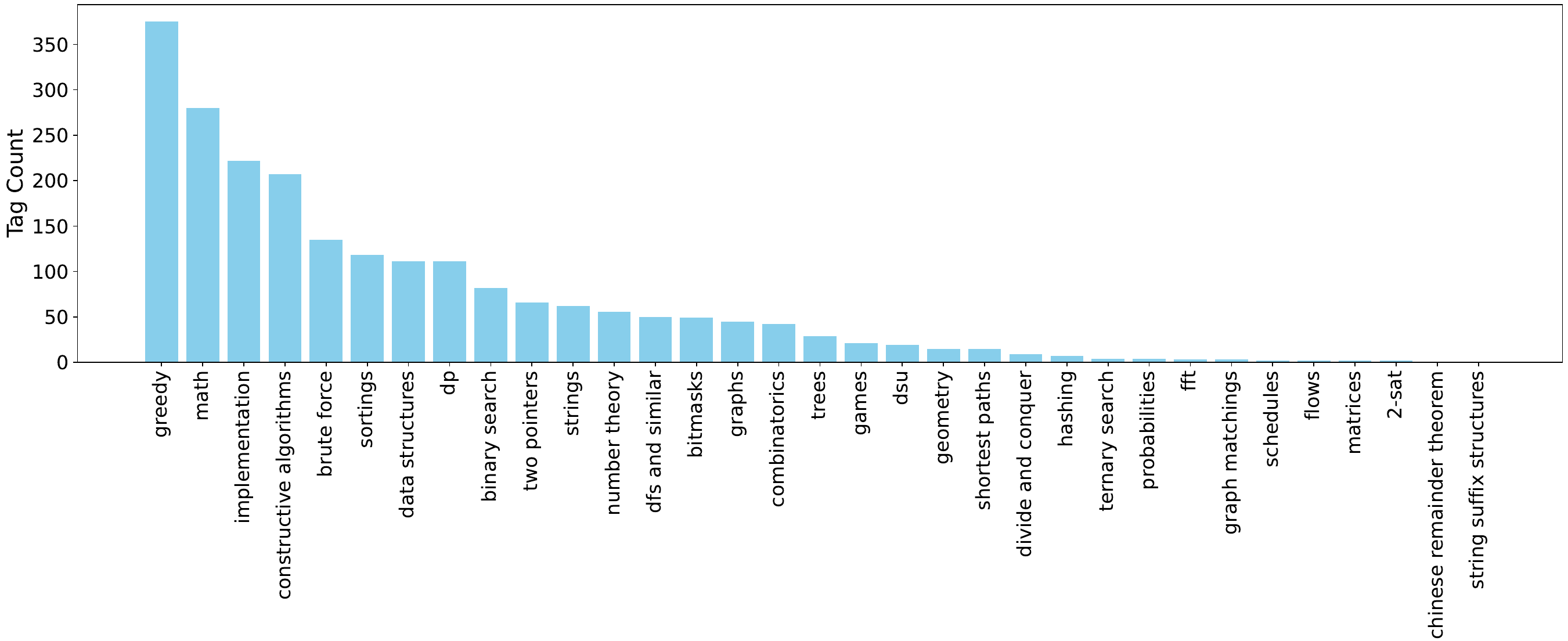}
        \caption{Distribution of CodeF post2021-9}
        \label{fig:sub3}
    \end{subfigure}
    \caption{Distribution of algorithms and data structure tags.}
    \label{fig3.1.2:Distribution_tags}
\end{figure}

According to the cutoff date of ChatGPT's training data - September 2021, we partitioned the dataset into CodeF pre2021-9 (comprising 805 problems) and CodeF post2021-9 (consisting of 718 problems). We got the raw data on the difficulty of the problems from the CodeForces website. The difficulty level of all problems is denoted in the format ``*xxx'' (ranging from *800-*3100\footnote{The difficulty level of which were sourced from the CodeForces programming website.}). In order to explore the rules of different difficulty problems, we classified the datasets according to the level of difficulty and extracted one hundred problems each from the Simple, Medium and Hard difficulty categories from both datasets, as depicted in Table \ref{tab:3.1datasetdifficulty}. 

To verify the reasonableness of the two parts of the data after we divided CodeF with the time of September 2021 as the node, we analyzed the average problem length and the average number of algorithm tags of the CodeF and the two parts of CodeF, as shown in Table \ref{tab:3.1.2datasetcondition}. We invited eight students with proficient knowledge of algorithms and data structures to examine the algorithm and data structures tags in our CodeF dataset, and eventually summarized 33 categories of tags. Meanwhile, in order to show the distribution of various types of algorithms and data structures on the problems of our dataset more intuitively, the distribution of tags was also summarized using statistical methods. The classification of all 33 algorithms and data structures tags in different parts of the CodeF dataset is shown in Figure \ref{fig3.1.2:Distribution_tags}.

Although our original intention of creating CodeF was to test the effectiveness of LLMs on programming problems of different vintages. After a series of rigorous processing during the construction of our dataset, in the end, CodeF dataset can be used as training data for code generation models, and the algorithms and data structures information in it is also can be applied to the fields of code understanding and code algorithm recommendation.

\subsection{Knowledge Library}\label{sec:3.2}
\begin{figure}[t]
  \centering
  \includegraphics[width=\linewidth]{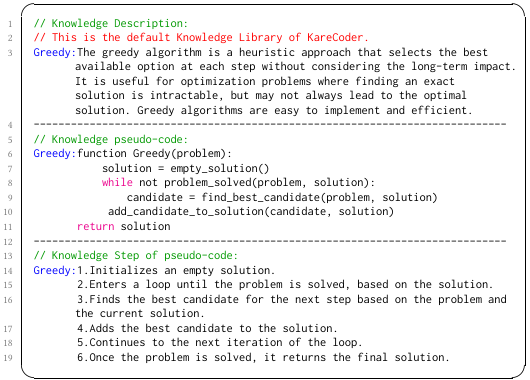}
  \caption{ Examples of different Knowledge Libraries. The corresponding Greedy algorithm is shown here.}
  \label{fig:3.2knowledgelibraryexample}
\end{figure}

\begin{figure*}[t]
  \centering
  \includegraphics[width=\linewidth]{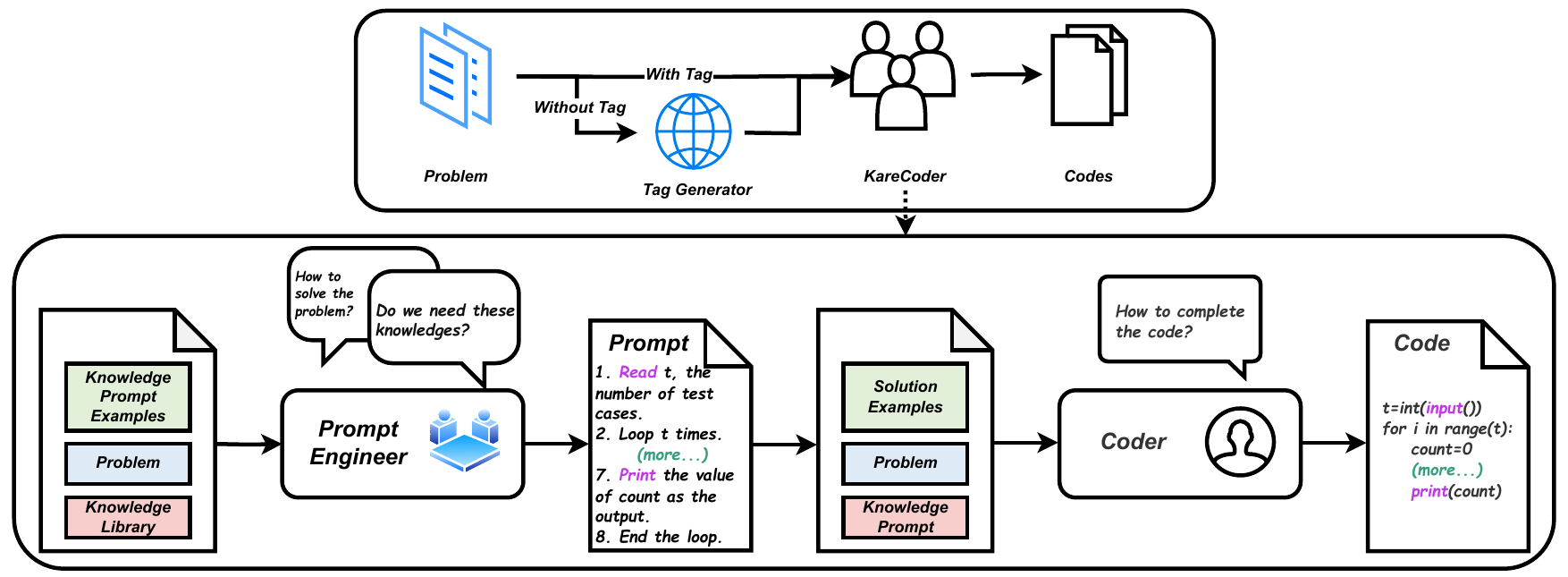}
  \caption{An overview of KareCoder. Given a programming problem, KareCoder matches the Knowledge Library appropriate to the problem and then generates a prompt for solving the problem. Finally, the prompt is used to guide the generation of code.}
  \label{fig:4.1KareCoder}
\end{figure*}

To enhance ChatGPT's proficiency in tackling novel programming problems, it is necessary to equip it with knowledge about algorithms and data structures. This knowledge serves as contextual reference information during the generation of prompts to address these problems. Specifically, we have conceptualized three formats of the Knowledge Library and investigated the effectiveness of these varying formats for prompt generation in RQ2 (see Section \ref{sec:6.2}).

We used resources such as ChatGPT and Google search engines to gather initial information and refer to the book Competitive Programmer's Handbook \cite{r34laaksonen2017competitive} for organization. Concurrently, we invited a group of individuals proficient in competitive programming knowledge. We manually refined and summarized the descriptions of each algorithm and data structure tag (these tags include all tags of CodeF problems). Our Knowledge Library of algorithms and data structures covers as much as possible the knowledge used in common programming problems. We invited seven employees and interns from companies in the programming industry to manually evaluate the correctness of the knowledge in our Knowledge Library. They evaluated all 33 pieces of knowledge and revised the final Knowledge Library to correctly explain each algorithm or data structure.

Our Knowledge Library is stored in the form of a dictionary, with each tag and its knowledge forming a one-to-one correspondence, i.e., \{``tag'': ``knowledge''\}. Ultimately, we integrated the information from the manually evaluated knowledge to build our Knowledge Library. Figure \ref{fig:3.2knowledgelibraryexample} provides a few examples from the Knowledge Library. The Knowledge Library we devised principally assumes the following three formats:

\begin{itemize}
\item \textbf{Knowledge Description:} This format provides a definitive overview of algorithms and data structures knowledge, encompassing descriptions of their properties and definitions. With this type of Knowledge Library, ChatGPT can comprehend the basic characteristics and applications of the corresponding algorithms and data structures, thereby generating appropriate problem-solving prompts. 
\item \textbf{Knowledge Pseudo-Code:} This Knowledge Library offers pseudo-code explanations of algorithms and data structures, emulating programming syntax to describe the steps and procedures involved in algorithms and data structures. By exploring this library, ChatGPT can understand the specific implementation of the corresponding programming knowledge, resulting in prompt skewed towards solution steps. 
\item \textbf{Knowledge Step of Pseudo-Code:} This form of Knowledge Library provides a textual step-by-step explanation of the pseudo-code associated with algorithms and data structures knowledge. With its assistance, ChatGPT can understand the steps and flow of programming knowledge from a natural language perspective. 
\end{itemize}

In our Knowledge Library, we offered descriptions, pseudo-code examples and steps of pseudo-code for 33 algorithms and data structure categories, such as greedy and dynamic programming. Essential to the KareCoder method, this content complements CodeF. Additionally, it can be integrated with the CodeF to train models for applications in code generation and code evaluation domains.

%% file: table/datasetdifficulty.tex
\begin{table}[]
\caption{ Subsets of Simple, Medium and Hard difficulties in CodeF Post2021-9 and CodeF Pre2021-9 datasets.}
\label{tab:3.1datasetdifficulty}
\centering
\begingroup
\setlength\tabcolsep{3pt}
\begin{tabular}{@{}cccc@{}}
\toprule
\textit{\textbf{Dataset}}                                                                      & \textit{\textbf{Difficulty}}                    & \textit{\textbf{Difficulty Range}} & \textit{\textbf{Problem Number}} \\ \midrule
                                                                                               & {\color[HTML]{000000} \textit{Simple}} & {\color[HTML]{000000} *800}              & {\color[HTML]{000000} 000-099}    \\
                                                                                               & {\color[HTML]{000000} \textit{Medium}} & {\color[HTML]{000000} *1300-*1600}       & {\color[HTML]{000000} 400-499}    \\
\multirow{-3}{*}{\textit{\textbf{\begin{tabular}[c]{@{}c@{}}CodeF\\ Post2021-9\end{tabular}}}}  & {\color[HTML]{000000} \textit{Hard}}   & {\color[HTML]{000000} *1900-*2500}       & {\color[HTML]{000000} 600-699}    \\ \hdashline 
                                                                                               & {\color[HTML]{000000} \textit{Simple}} & {\color[HTML]{000000} *800}              & {\color[HTML]{000000} 000-099}    \\
                                                                                               & {\color[HTML]{000000} \textit{Medium}} & {\color[HTML]{000000} *1300-*1600}       & {\color[HTML]{000000} 400-499}    \\
\multirow{-3}{*}{\textit{\textbf{\begin{tabular}[c]{@{}c@{}}CodeF\\ Pre2021-9\end{tabular}}}} & {\color[HTML]{000000} \textit{Hard}}   & {\color[HTML]{000000} *1900-*2200}       & {\color[HTML]{000000} 640-739}    \\ \bottomrule
\end{tabular}
\endgroup
\end{table}

%% file: table/dataset-condition-new.tex
\begin{table}[]

\caption{Descriptive statistics of CodeF.}
\label{tab:3.1.2datasetcondition}
\centering
\begin{tabular}{@{}
>{\columncolor[HTML]{FFFFFF}}l 
>{\columncolor[HTML]{FFFFFF}}c 
>{\columncolor[HTML]{FFFFFF}}c @{}}
\toprule
{\color[HTML]{1F2328} \textit{\textbf{Dataset}}}          & \multicolumn{1}{l}{\cellcolor[HTML]{FFFFFF}{\color[HTML]{1F2328} \textit{\textbf{\begin{tabular}[c]{@{}l@{}}Average question\\ token\end{tabular}}}}} & \multicolumn{1}{l}{\cellcolor[HTML]{FFFFFF}{\color[HTML]{1F2328} \textit{\textbf{\begin{tabular}[c]{@{}l@{}}Average algorithm\\ tags number\end{tabular}}}}} \\ \midrule
{\color[HTML]{1F2328} \textit{\textbf{CodeF}}}            & {\color[HTML]{000000} 2047.1}                                                                                                                         & {\color[HTML]{000000} 3.1}                                                                                                                                   \\
{\color[HTML]{1F2328} \textit{\textbf{CodeF pre2021-9}}}  & {\color[HTML]{000000} 2018.7}                                                                                                                         & {\color[HTML]{000000} 3.2}                                                                                                                                   \\
{\color[HTML]{1F2328} \textit{\textbf{CodeF post2021-9}}} & {\color[HTML]{000000} 2079.0}                                                                                                                         & {\color[HTML]{000000} 3.0}                                                                                                                                   \\ \bottomrule
\end{tabular}
\end{table}

%% file: sec/4.approach.tex
\section{Approach}\label{sec:4}
In this section, we present a new method: KareCoder. We begin with an overview of KareCoder and present it in two stages. 

\subsection{Overview}\label{sec:4.1}

Code generation aims to help solve programming problems. We postulate that the thought process of LLMs parallels that of human programmers, which can be bifurcated into two stages when tackling programming problems: initially, comprehending the problem and learning or recalling knowledge that may be applied to its solution, followed by formulating a preliminary plan; secondly, executing code writing based on this plan. 

While a majority of problems on programming platforms come with algorithmic or data structure tags, there exists a subset of problems from non-programming platforms or datasets, such as APPS, lacking these algorithmic tags. To enhance the adaptability of KareCoder, we developed a tag generator, drawing from ChatGPT and Prompt technologies and based on our Knowledge Library. This generator can formulate or rectify tags, especially for problems that either lack them or have tags misaligned with our Knowledge Library.  We evaluated the tag generator’s accuracy with Cohen’s Kappa and manual evaluation methods. Leveraging this tag generator, we ensure a seamless integration of each programming problem into KareCoder's processing framework. Consequently, as depicted in Figure \ref{fig:4.1KareCoder}, we connected a tag generator in front of KareCoder and partitioned the workflow of KareCoder into two stages, Prompt Engineering and Coding, adhering to a sequential approach to the task:

\begin{itemize}
\item In \textbf{Prompt Engineering Stage}, we designate ChatGPT as a prompt engineer, accountable for comprehending the problem and learning as well as reviewing the algorithms and data structures knowledge pertinent to resolving the problem, subsequently generating a problem-solving prompt. 
\item In \textbf{Coding Stage}, we assign ChatGPT the role of a coder, tasked with perusing the problem and implementing this solution in a programming language, guided by the prompt provided by the prompt engineer. 
\end{itemize}

\begin{figure}[t]
  \centering
  \includegraphics[width=\linewidth]{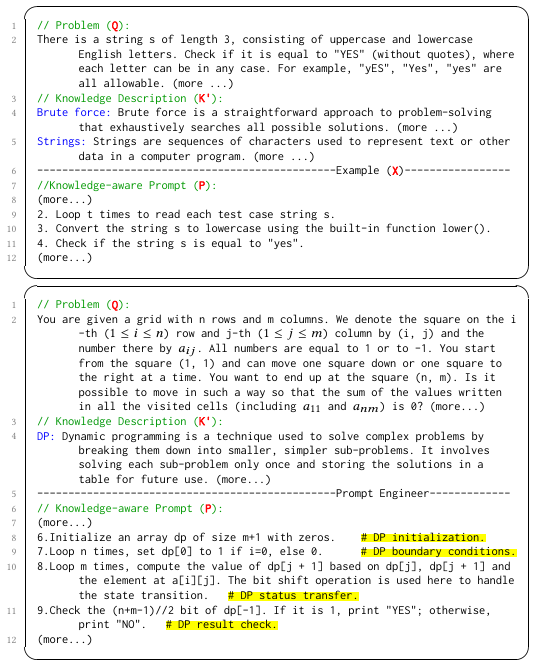}
  \caption{Illustration of the input and output of Prompt Engineering Stage. The input includes a sample, a new problem and its corresponding Knowledge Library, the output is the knowledge-aware prompt for the new problem.}
  \label{fig:4.2promptengineer}
\end{figure}

\subsection{Prompt Engineering Stage}\label{sec:4.2}
In Prompt Engineering Stage, we regard ChatGPT as a prompt engineer. We anticipate that ChatGPT will acquire the relevant knowledge needed for resolving programming problems, subsequently, based on its pre-existing knowledge complemented by newly recalled information, it will develop a prompt for addressing the problem. Given ChatGPT's input window constraints, inputting the entire Knowledge Library isn't viable. Therefore, we update and customize the knowledge for each problem by performing absolute value matching via dictionaries to associate the tags of the problem with the tags in the Knowledge Library. The working principle of dictionary matching is as follows: The corresponding relationships between the dataset and the Knowledge Library are <problem, tag> and <tag, knowledge>, respectively. We can achieve matching through the tag in both of them.

\begin{figure}[t]
  \centering
  \includegraphics[width=\linewidth]{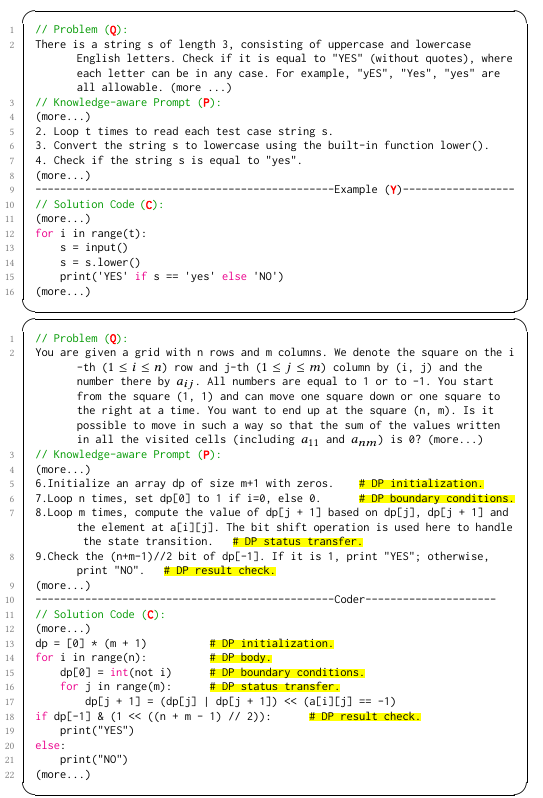}
  \caption{Illustration of the input and output of Coding Stage. The input includes a sample, a new problem and their corresponding prompts for generating code, the output is the solution code for the new problem.}
  \label{fig:4.3coder}
\end{figure}

As depicted in Figure \ref{fig:4.2promptengineer}, we manually constructed a one-shot example ($X$) that generates a knowledge-aware prompt ($P$) based on the problem ($Q$) and the knowledge description ($K'$) in our Knowledge Library ($K$). The one-shot example ($X$) we developed was curated by eight students involved in the review of CodeF's algorithm and data structure tags. Following the review, we identified four representative examples for experimentation, ultimately selecting one as the one-shot example. Subsequently, we concatenated the new problem with a DP (dynamic programming) knowledge description derived from the Knowledge Library. Relying on the one-shot example we provided and the DP algorithms pertinent to the new problem, we crafted a step-by-step problem-solving, knowledge-aware prompt. Full details are available on our GitHub\footnote{\url{https://github.com/CodeGeneration3/KareCoder}}.

Let $Q$ represent the problem, $P$ represent the knowledge-aware prompt, $K$ signify the Knowledge Library, $K'$ signify the knowledge description, $X$ denote a set of examples utilized to generate prompts based on the problem and the knowledge description, i.e., $X = \left\{ \left\langle Q_{x}, K'_{x}, P_{x} \right\rangle \right\}_{x=1}^{n}$. Consequently, the knowledge-aware prompt generation problem can be modelled as:
\begin{equation}
f\left ( P\mid Q, K, X\right )\triangleq f\left ( P\mid Q, K',X \right )\cdot m\left ( K'\mid Q, K \right )  
\end{equation}

Herein, $f$ represents the generative tasks completed using ChatGPT, $m$ signifies the dictionary-matching tasks of knowledge. 
\subsection{Coding Stage}\label{sec:4.3}
During Coding Stage, we conceptualize ChatGPT as a Coder, with the expectation that it can integrate the problem and the knowledge-aware prompt to generate code that resolves the problem. The objective of this stage is to transform a step-by-step prompt, represented in natural language, into an executable program. As illustrated in Figure \ref{fig:4.3coder}, this diagram schematically portrays the Coding Stage. Full details are available on our GitHub.

Let $Q$ represent the problem, $P$ stand for the knowledge-aware prompt, $C$ symbolize the generated code, and $Y$ correspond to a set of examples pertaining to code generation, which are predicated on the problem and the associated prompt, i.e., $Y = \left\{ \left\langle Q_{y}, P_{y}, C_{y} \right\rangle \right\}_{y=1}^{n}$. The one-shot example ($Y$) was summarized using the same approach as one-shot example ($X$). Therefore, the task of code generation based on the prompt can be mathematically formulated as :
\begin{equation}
f\left ( C\mid Q, P, Y \right ) 
\end{equation}

Subsequently, the comprehensive KareCoder approach can be mathematically represented as:
\begin{equation}
f(C\mid Q, K)\triangleq \underbrace{f(C\mid Q, P, Y)\cdot}_{\text{Generate-Code}}\underbrace{f(P\mid\mathrm{Q, K', X})\cdot}_{\text{Generate-Prompt}}\underbrace{m(K'\mid Q, K)}_{\text{Match-Knowledge}}
\end{equation}

%% file: sec/5.environmentsetup.tex
\section{Environment Setup}\label{sec:5}
\subsection{Reserach Questions}\label{sec:5.1}
In this research, we introduce KareCoder that supplements ChatGPT with a Knowledge Library to guide its operation. To evaluate the efficacy of KareCoder and to analyze the associated influencing factors, we explore the following research questions (RQs):

\textbf{RQ1: How does the KareCoder perform compared to other baselines?} The objective of this RQ is to ascertain whether KareCoder can produce programs of greater accuracy than other methods. We conducted comparisons using the CodeF post2021-9 dataset as well as subsets of the dataset delineated by levels of difficulty - Simple, Medium and Hard. 

\textbf{RQ2: What is the better choice for the Knowledge Library?} Determining how to enable ChatGPT to better utilize programming knowledge without constraining its innate reasoning abilities is a challenge we need to address. Consequently, we have designed three distinctive forms of the Knowledge Library for evaluation. 

\textbf{RQ3: Will different shot times and ChatGPT Settings affect the KareCoder performance?} Within this RQ, we examined the impact of employing k-shot examples on the effectiveness of KareCoder and whether variations in parameter settings affect KareCoder's performance under a 1-shot example condition. 

\textbf{RQ4: Is KareCoder only valid because of the datasets we used?} This can be a confusing issue for many people. Hence, we also tested our methods on the first 500 problems from the APPS test set to demonstrate that the efficacy of KareCoder is not contingent upon a specific dataset.

\subsection{Benchmarks}\label{sec:5.2}

\textbf{CodeF Dataset:} CodeF is a Python task dataset that we developed. These problems are sourced from the CodeForces website and range from January 2020 to April 2023, encompassing a total of 1523 problems. Of these, 805 problems are from the period prior to September 2021 and 718 problems are from the subsequent period. Our experiments are primarily focused on the problems posed after September 2021. From these problems, we extracted three subsets, namely Simple, Medium and Hard, based on the level of difficulty.

\textbf{APPS Dataset:} We used the APPS dataset as an auxiliary validation source in our experiments. The APPS dataset is derived from various programming competition websites, including CodeForces, LeetCode, CodeChef, etc. The training set comprises 5,000 problems, while the test set also contains 5,000 problems. In our experiments, we selected the first 500 problems of the test set for our tests.

\subsection{Comparison Baselines}\label{sec:5.3}
For the purposes of comparison, we selected ChatGPT, two recently proposed methods for code generation, specifically, Self-planning \cite{r14jiang2023self} and SCOT \cite{r31li2023enabling}, and SCOT\&KareCoder which combines the Structured COT and Knowledge-aware method as baselines. Simultaneously, to further investigate the capabilities of each method, we crafted a system prompt tailored for the four baselines methods with the aim of enhancing the quality of the generation results. 

\textbf{ChatGPT} \cite{r7OpenAI2022} is an LLM proposed by OpenAI, it exhibits characteristics similar to a question-and-answer system, with a powerful code-writing ability. It exhibits exceptional performance in few-shot cases. For our experiments, we utilized OpenAI's API to call ChatGPT with the specific model ``gpt-3.5-turbo-0613''.

\textbf{Plan} is our implementation inspired by Self-planning \cite{r14jiang2023self}. The Self-planning strategy encompasses two phases: planning phase and implementation phase. Drawing from its idea, we fashioned a two-phase process: initially, we allow ChatGPT to generate a step-by-step prompt for problem-solving, following which, the code of the problem is generated under the guidance of the prompt. 

\textbf{SCOT} \cite{r31li2023enabling} aimed at investigating how to access the coding mindset of LLMs in the context of code generation. SCOT first generates a Structured COT, delineating the solution process in programming logic, then transmutes this Structured COT into a program using a specific programming language. SCOT specifies a temperature setting of 0.8 and a Top\_P value of 0.95, we adhered to these parameters in our experiments. 

\textbf{SCOT\&KareCoder} is a way to combine KareCoder's knowledge-aware prompt with SCOT's Structured COT. This baseline differs from KareCoder in that instead of generating a step-by-step prompt similar to the Plan approach in the Prompt Engineering Stage, we generate a Knowledge-aware Structured COT to guide the code generation.

\subsection{Metrics}\label{sec:5.4}
To evaluate the accuracy of the generated code, we used Pass@k as an evaluation metric, as shown in the formula below. For each problem, we generate n $\geq$ k copies of code and calculate the number of correct samples that pass the test sample, c (where c $\geq$ n). In our experiments, we generated 5 copies of code for each problem and evaluated our approach using Pass@1, Pass@3 and Pass@5. Notably, Pass@1 is especially important as it aligns more with practical application needs.

\begin{equation}
Pass@k = \mathop{\mathbb{E}}\limits_{Problems}\begin{bmatrix}1-\frac{\begin{pmatrix}n-c\\k\end{pmatrix}}{\begin{pmatrix}n\\k\end{pmatrix}}\end{bmatrix}.
\end{equation}

%% file: sec/6.resultandanalysis.tex
\section{Result and Analysis}\label{6}
\input{table/rq1simplemediumhard}

In conducting the experiments for these research questions, we adhere to a default configuration of a one-shot setting, with the temperature parameter set to 1 and the Top\_p parameter also set to 1. Should there be alterations to this configuration, we will explicitly specify them. The comparative rule utilized in our study involves contrasting KareCoder (marked in blue in the tables) with the other best methods (marked in yellow in the tables).
\subsection{RQ1: Performance Comparison}\label{sec：6.1}
\textbf{1) Evaluation on Simple, Medium and Hard difficulties of CodeF Post2021-9.} Within RQ1, we examined the effectiveness of the knowledge introduced by KareCoder on three difficulty levels (Simple, Medium and Hard) within the CodeF post2021-9 dataset through a series of comparative experiments, which are depicted in Table \ref{tab:6.1.1rq1simplemeidumhard}. We assessed the outcomes of direct code generation by ChatGPT with 1-shot, as well as Plan, SCOT, SCOT\&KareCoder and KareCoder under 1-shot and 3-shot conditions, respectively. The analyses of these results are as follows:
\begin{itemize}
\item KareCoder approach markedly excelled beyond other approaches in terms of the Pass@1 performance metric. This enhancement is largely ascribed to the creation of a knowledge-aware prompt, individually tailored to each programming problem. Such prompts not only directed the code generation process but also markedly bolstered the code's correctness, aligning with our objective of augmenting the Pass@1 performance index.
\item Nevertheless, in Pass@3 and Pass@5 metrics, KareCoder still outperformed other methods, but it didn’t achieve the significant gains that direct code generation with ChatGPT did. This results likely arises from the inherent risks associated with the intermediate step of prompt or Structured COT generation. Though designed to facilitate code production, inaccuracies in these generated elements have the potential to compromise the precision of all subsequent code generation pertaining to the specific problem at hand.
\item In our study, KareCoder outshone in 1-shot Pass@k metrics, except Hard level, with over 20\% advantages in Pass@3 and Pass@5 in simple and medium level. In addition, KareCoder's in the 3-shot experiment did not result in an absolute advantage. However, multiple shots may lead to waste of resources caused by long input length. Thus, we favor fewer-shot techniques, a preference supported by our RQ3 experiment assessing KareCoder’s efficiency.
\end{itemize}
\textbf{2) Evaluation on the CodeF Post2021-9.} Further tests on the CodeF Post2021-9 dataset confirm KareCoder's efficacy. As Table \ref{tab:6.1.2rq1codefpost2021-9} illustrates, on the Pass@1 metric, KareCoder and the Plan method achieve 15.9\% and 14.2\% respectively, ranking first and second, far more than the results generated directly by ChatGPT; conversely, in terms of the Pass@3 and Pass@5 metrics, ChatGPT's direct generation remains the most effective. The rationale behind this phenomenon has been comprehensively discussed in the previous part of this RQ. Compared with the Plan, SCOT and SCOT\&KareCoder methods, KareCoder attains a relative improvement of 12.0\%, 14.5\% and 16.4\% on the Pass@1, Pass@3 and Pass@5 metrics, respectively. 
\input{table/rq1codefpost2021-9}

\find{
\textbf{Answer to RQ1:} Experiments on CodeF post2021-9 and its subsets—Simple, Medium and Hard show that KareCoder considerably enhances ChatGPT's performance on the Pass@1 metric. Nevertheless, due to constraints arising from the intermediate process, KareCoder does not outperform the direct usage of ChatGPT on the Pass@3 and Pass@5 metrics but maintains advantage over other methods when evaluated on the Pass@3 and Pass@5 metrics.}

\input{table/rq2knowledgelibrary}

\input{table/rq3k-shot}

\input{table/rq3tem_top}

\subsection{RQ2: Effectiveness of Knowledge Type}\label{sec:6.2}
\textbf{1) Library Variants.} The Knowledge Library is a critical component of KareCoder. This research proposes a Knowledge Library dedicated to algorithms and data structures, specifically designed for programming problems. This innovation aims to addresses ChatGPT's ``knowledge forgetting'' during solution planning, specifically the loss of necessary programming knowledge. For this research question, we investigated various Knowledge Library alternatives and contrasted different design possibilities. We built three distinct Knowledge Libraries, namely: Knowledge Pseudo-code, Knowledge Steps of Pseudo-code and the Knowledge Description employed by KareCoder. Further details regarding these diverse Knowledge Libraries can be found in Section \ref{sec:3.2}.

\textbf{2) Results.} As demonstrated in Table \ref{tab:6.2rq2knowledgelibrary}, among the three Knowledge Librarys, the Knowledge Description yielded the most superior performance, achieving the highest scores across all difficulty levels on the Pass@1, Pass@3 and Pass@5 metrics. Notably, its relative improvement reached 15.1\%, 16.7\% and 13.8\% on the Pass@1 metric. Moreover, KareCoder utilizing the Knowledge Description consistently outperforms the Plan method without a Knowledge Library. These results suggest that LLMs exhibit improved performance in handling novel problems when suitable external information is integrated. Conversely, the integration of unsuitable external information could potentially yield adverse effects. 

\find{
\textbf{Answer to RQ2:} Based on the current analysis, the Knowledge Description of the Knowledge Library has exhibited the most promising results. Nonetheless, we conjecture that the performance of KareCoder could be further enhanced by constructing more refined Knowledge Libraries and implementing superior knowledge-importing techniques in the future.}

\subsection{RQ3: Influence of Shot Times and Settings}\label{sec:6.3}
In this RQ, we scrutinize the performance of KareCoder in relation to varying shot times and parameter settings for ChatGPT, thereby performing a stability analysis of the KareCoder method. 

\textbf{1) Impact of Shot Times.} As indicated in Table \ref{tab:6.3.1rq3k-shot}, the results derived from 1-shot, 2-shot and 3-shot experiments show that, KareCoder's effectiveness does not significantly fluctuate in response to variations in shot times. When implementing ChatGPT, we must also account for constraints related to maximum input length. While the ``gpt-3.5-turbo-16k-0613'' model can accommodate greater input length, it is crucial to strike a balance between shot times and performance, taking into account considerations related to computational resources and associated costs. Taking these factors into account, it is reasonable to conclude that the 1-shot setting adequately caters to the vast majority of the method's requirements.

\textbf{2) Impact of ChatGPT Settings.} Within the RQ, we investigate the degree to which various parameter settings, such as Temperature and Top\_P, impact the efficacy of KareCoder. As illustrated in Table \ref{tab:6.3.2rq3tem_top_P}, KareCoder performs optimally when both Temperature and Top\_P are designated as 1. This may be attributed to the fact that we furnished ChatGPT with an external Knowledge Library and we did not impose stringent restrictions on the implementation of algorithms and data structures within the Knowledge Library. Instead, we intended the Knowledge Library to serve as a guide, while KareCoder necessitated a significant degree of randomness. Establishing Temperature=1 and Top\_P=1 endows ChatGPT with greater randomness, thereby circumventing the generation of chained error code owing to errors encountered during the production of knowledge-aware prompt.

\find{
\textbf{Answer to RQ3:} Varying Shot times, 1-shot, 2-shot and 3-shot, each setting exhibits unique benefits and limitations. Nonetheless, when considering computational resource constraints, the 1-shot setting is already sufficient to meet the accuracy demands of code generation. As for the parameter configurations of ChatGPT, we select the default settings, specifically, Temperature=1 and Top\_P=1, to preserve the randomness of the outputs generated by ChatGPT  to enhance the effectiveness of KareCoder.}

\subsection{RQ4: Impact of Different Datasets}\label{sec:6.4}

In an effort to comprehensively assess the efficacy of KareCoder, we selected the first 500 problems from the open-source dataset, the APPS test set and contrasted KareCoder's performance with other methods on these problems. As indicated in Table \ref{tab:6.4rq4apps}, five methods were employed: direct code generation using ChatGPT, Plan, SCOT, SCOT\&KareCoder and KareCoder, the performance of the other four methods has not been able to surpass ChatGPT, the phenomenon that has been verified in previous research \cite{r15li2023think}. In light of the experimental results, we hypothesize that this may be attributed to overlapping data within APPS that is present in the ChatGPT training set. During the training process of ChatGPT, the massive number of natural language-code pairs (<Text, Code>) serve as training data. Consequently, ChatGPT can resolve these problems without the prerequisite for problem planning. 

In a comprehensive evaluation, KareCoder's performance is approximately on par with the best of other methods and does not manifest a marked superiority. This condition potentially be ascribed to ChatGPT's prior exposure to open-source problems in APPS, resulting in the additional knowledge failing to deliver anticipated guidance. Nonetheless, in the experiments contrasting these methods, KareCoder exhibits superior performance across the Pass@k metric compared to the Plan method, which employs the same step-by-step prompt. Furthermore, the SCOT\&KareCoder, which incorporates external knowledge, exhibits a slightly lower performance than SCOT alone on the Pass@1 metric but significantly surpasses SCOT on the Pass@3 and Pass@5 metrics. From this phenomenon, we can still discover the role of incorporating knowledge into code generation through a two-by-two comparison of Plan and KareCoder, SCOT and SCOT\&KareCoder.
\input{table/rq4apps}

\find{
\textbf{Answer to RQ4:} Through experimentation on the first 500 problems drawn from the APPS test set, the outcomes directly generated by ChatGPT markedly exceed those of the other four methods. Despite this, KareCoder retained advantages over the other three methods, KareCoder's performance is approximately on par with the best of other methods.}

%% file: table/rq1simplemediumhard.tex
\begin{table*}[]
\centering
\caption{Comparative evaluation of KareCoder and other methods on Simple, Medium and Hard difficulties of CodeF Post2021-9.}
\vspace{-0.5em}
\label{tab:6.1.1rq1simplemeidumhard}
\begingroup
\setlength\tabcolsep{4.2pt}
\begin{tabular}{@{}lccccccccc@{}}
\toprule
                                                                                           & \multicolumn{3}{c}{\textit{\textbf{Simple}}}                                                 & \multicolumn{3}{c}{\textit{\textbf{Medium}}}                                                   & \multicolumn{3}{c}{\textit{\textbf{Hard}}}                                                    \\ \cmidrule(l){2-4} \cmidrule(l){5-7} \cmidrule(l){8-10} 
\multirow{-2}{*}{\textit{\textbf{Method}}}                                                 & \textit{\textbf{Pass@1}}     & \textit{\textbf{Pass@3}}      & \textit{\textbf{Pass@5}}     & \textit{\textbf{Pass@1}}       & \textit{\textbf{Pass@3}}      & \textit{\textbf{Pass@5}}     & \textit{\textbf{Pass@1}}      & \textit{\textbf{Pass@3}}      & \textit{\textbf{Pass@5}}     \\ \midrule
\textit{\textbf{ChatGPT}}                                                                  & 26.1                         & 39.8                          & 46.2                          & 7.0                            & 15.0                          & 19.5                          & 6.0                           & 10.4                          & 12.4                          \\ \midrule
\textit{\textbf{ChatGPT+Plan}}                                                             & \coloryellow $28.5$                         & \coloryellow $31.8$                          & \coloryellow$33.6$                          & 5.1                            & 7.6                           & 8.4                           & 7.3                           & 8.6                           & 9.0                          \\
\textit{\textbf{ChatGPT+SCOT}}                                                              & 19.1                         & 24.3                          & 25.2                          & 6.4                            & \coloryellow$11.3$                          & \coloryellow$13.1$                          & \coloryellow{\textbf{7.7}}                  & \coloryellow {\textbf{11.6}}                          & \coloryellow $12.4$                          \\
\textit{\textbf{\begin{tabular}[c]{@{}l@{}}ChatGPT+SCOT\&KareCoder\end{tabular}}} & 24.7                         & 25.0                          & 25.0                          & \coloryellow $7.7$                            & 9.2                         & 9.6                          & 7.3                           & 8.0                           & 8.0                          \\
\textit{\textbf{ChatGPT+KareCoder}}                                                        & \colorblue{\textbf{30.5}}                & \colorblue{\textbf{38.8}}                 & \colorblue{\textbf{41.8}}                 & \colorblue{\textbf{10.5}}                  & \colorblue{\textbf{14.4}}                 & \colorblue{\textbf{16.4}}                 & \colorblue $7.4$                           & \colorblue $11.1$                & \colorblue{\textbf{12.5}}                 \\ \hdashline
\textit{\textbf{Relative Improvement}}                                                     & {\color[HTML]{FF0000} 7.0\% $\uparrow$} & {\color[HTML]{FF0000} 22.0\% $\uparrow$} & {\color[HTML]{FF0000} 24.4\% $\uparrow$} & {\color[HTML]{FF0000} 36.4\% $\uparrow$}  & {\color[HTML]{FF0000} 27.4\% $\uparrow$} & {\color[HTML]{FF0000} 25.2\% $\uparrow$} & {\color[HTML]{0500FF} -3.9\% $\downarrow$} & {\color[HTML]{0500FF} -4.3\% $\downarrow$}  & {\color[HTML]{FF0000} 0.8\% $\uparrow$}  \\ \midrule
\textit{\textbf{ChatGPT+Plan(3-shot)}}                                                             & 29.7                         & \coloryellow $31.9$                          & \coloryellow$32.4$                          & 7.1                            & \coloryellow{\textbf{12.5}}                           & \coloryellow{\textbf{15.8}}                           & 6.5                           & 7.9                           & 8.5                          \\
\textit{\textbf{ChatGPT+SCOT(3-shot)}}                                                      & \coloryellow$29.9$                         & 30.0                          & 30.0                          & \coloryellow{\textbf{10.8}}                  & 12.2                 & 12.5                 & 7.7                           & 9.4                           & 9.8                          \\
\textit{\textbf{\begin{tabular}[c]{@{}l@{}}ChatGPT+SCOT\&KareCoder(3-shot)\end{tabular}}} & 29.7                         & 30.0                          & 30.0                          & 9.1                            & 10.5                         & 10.9                          & \coloryellow$8.8$                           & \coloryellow$9.8$                           & \coloryellow$9.9$                          \\
\textit{\textbf{\begin{tabular}[c]{@{}l@{}}ChatGPT+KareCoder(3-shot)\end{tabular}}}     & \colorblue{\textbf{31.5}}                & \colorblue{\textbf{34.1}}                 & \colorblue{\textbf{34.9}}                & \colorblue $9.3$                            & \colorblue$10.9$                          & \colorblue$11.5$                          & \colorblue{\textbf{9.6}}                  & \colorblue{\textbf{11.7}}                 & \colorblue{\textbf{12.5}}                 \\ \hdashline
\textit{\textbf{Relative Improvement}}                                                     & {\color[HTML]{FF0000} 5.4\% $\uparrow$} & {\color[HTML]{FF0000} 6.9\% $\uparrow$} & {\color[HTML]{FF0000} 7.7\% $\uparrow$} & {\color[HTML]{0500FF} -13.9\% $\downarrow$} & {\color[HTML]{0500FF} -12.8\% $\downarrow$} & {\color[HTML]{0500FF} -27.2\% $\downarrow$} & {\color[HTML]{FF0000} 9.1\% $\uparrow$}  & {\color[HTML]{FF0000} 19.4\% $\uparrow$} & {\color[HTML]{FF0000} 26.3\% $\uparrow$} \\ \bottomrule
\end{tabular}
\endgroup
\end{table*}

%% file: table/rq1codefpost2021-9.tex
\begin{table}[]
\caption{ Comparative evaluation of KareCoder with other methods on the CodeF post2021-9. }
\label{tab:6.1.2rq1codefpost2021-9}
\centering
\begin{tabular}{@{}lccc@{}}
\toprule
\textit{\textbf{Method}}               & \textit{\textbf{Pass@1}}      & \textit{\textbf{Pass@3}}      & \textit{\textbf{Pass@5}}     \\ \midrule
\textit{\textbf{ChatGPT}}              & 12.9                          & 22.4                          & 26.9                          \\ \midrule
\textit{\textbf{ChatGPT+Plan}}         & \coloryellow $14.2$                          & \coloryellow $17.3$                          & \coloryellow$18.3$                          \\
\textit{\textbf{ChatGPT+SCOT}}          & 11.2                          & 14.2                          & 15.0                          \\
\textit{\textbf{ChatGPT+SCOT\&KareCoder}}          & 14.0                          & 15.2                          & 15.5                          \\
\textit{\textbf{ChatGPT+KareCoder}}    & \colorblue{\textbf{15.9}}                & \colorblue{\textbf{19.8}}                 & \colorblue {\textbf{21.3}}                 \\  \hdashline 
\textit{\textbf{Relative Improvement}} & {\color[HTML]{FF0000} 12.0\% $\uparrow$} & {\color[HTML]{FF0000} 14.5\% $\uparrow$} & {\color[HTML]{FF0000} 16.4\% $\uparrow$} \\ \bottomrule
\end{tabular}
\end{table}

%% file: table/rq2knowledgelibrary.tex
\begin{table*}[]
\caption{Performance of KareCoder utilizing different Knowledge Libraries. }
\label{tab:6.2rq2knowledgelibrary}
\centering
\begingroup
\setlength\tabcolsep{4.5pt}
\begin{tabular}{@{}lccccccccc@{}}
\toprule
                                                      & \multicolumn{3}{c}{\textit{\textbf{Simple}}}                                                  & \multicolumn{3}{c}{\textit{\textbf{Medium}}}                                                  & \multicolumn{3}{c}{\textit{\textbf{Hard}}}                                                  \\ \cmidrule(l){2-4} \cmidrule(l){5-7} \cmidrule(l){8-10}
\multirow{-2}{*}{\textit{\textbf{Knowledge Library}}} & \textit{\textbf{Pass@1}}      & \textit{\textbf{Pass@3}}      & \textit{\textbf{Pass@5}}     & \textit{\textbf{Pass@1}}      & \textit{\textbf{Pass@3}}      & \textit{\textbf{Pass@5}}     & \textit{\textbf{Pass@1}}      & \textit{\textbf{Pass@3}}      & \textit{\textbf{Pass@5}}   \\ \midrule
\textit{\textbf{Knowledge Description}}               & \colorblue{\textbf{30.5}}                 & \colorblue{\textbf{38.8}}                 & \colorblue{\textbf{41.8}}                 & \colorblue{\textbf{10.5}}                 & \colorblue{\textbf{14.4}}                 & \colorblue{\textbf{16.4}}                 & \colorblue{\textbf{7.4}}                  & \colorblue{\textbf{11.1}}                 & \colorblue{\textbf{12.5}}               \\
\textit{\textbf{Knowledge Pseudo-code}}               & 23.5                          & 27.3                          & 28.7                          & \coloryellow $9.0$                           & \coloryellow $12.7$                          & \coloryellow $14.2$                          & \coloryellow $6.5$                           & \coloryellow $8.0$                           & 8.6                         \\
\textit{\textbf{Knowledge Step of Pseudo-code}}       & \coloryellow $26.5$                          & \coloryellow $29.3$                          & \coloryellow $29.8$                          & 7.4                           & 10.7                          & 12.0                          & 5.9                           & 7.9                           & \coloryellow $8.8$                        \\ \hdashline 
\textit{\textbf{Relative Improvement}}                & {\color[HTML]{FF0000} 15.1\% $\uparrow$} & {\color[HTML]{FF0000} 32.4\% $\uparrow$} & {\color[HTML]{FF0000} 40.3\% $\uparrow$} & {\color[HTML]{FF0000} 16.7\% $\uparrow$} & {\color[HTML]{FF0000} 13.4\% $\uparrow$} & {\color[HTML]{FF0000} 15.5\% $\uparrow$} & {\color[HTML]{FF0000} 13.8\% $\uparrow$} & {\color[HTML]{FF0000} 38.8\% $\uparrow$} & {\color[HTML]{FF0000} 42.0\% $\uparrow$} \\ \bottomrule
\end{tabular}
\endgroup
\end{table*}

%% file: table/rq3k-shot.tex
\begin{table*}[]
\caption{Comparative evaluation of KareCoder with different shot times on various difficulties of CodeF post2021-9.}
\label{tab:6.3.1rq3k-shot}
\centering
\begin{tabular}{@{}lccccccccc@{}}
\toprule
                                               & \multicolumn{3}{c}{\textit{\textbf{Simple}}}                                                  & \multicolumn{3}{c}{\textit{\textbf{Medium}}}                                                & \multicolumn{3}{c}{\textit{\textbf{Hard}}}                                                 \\ \cmidrule(l){2-4} \cmidrule(l){5-7} \cmidrule(l){8-10}
\multirow{-2}{*}{\textit{\textbf{Shot Times}}} & \textit{\textbf{Pass@1}}      & \textit{\textbf{Pass@3}}      & \textit{\textbf{Pass@5}}     & \textit{\textbf{Pass@1}}     & \textit{\textbf{Pass@3}}      & \textit{\textbf{Pass@5}}    & \textit{\textbf{Pass@1}}       & \textit{\textbf{Pass@3}}   & \textit{\textbf{Pass@5}}    \\ \midrule
\textit{\textbf{1-shot}}                       & \colorblue$30.5$                          & \colorblue{\textbf{38.8}}                 & \colorblue{\textbf{41.8}}                 & \colorblue$10.5$                         & \colorblue$14.4$                          & \colorblue{16.4}                & \colorblue{7.4}                            & \colorblue{11.1}              & \colorblue{\textbf{12.5}}                \\
\textit{\textbf{2-shot}}                       & 30.7                          & \coloryellow$34.2$                          & \coloryellow$36.1$                          & \coloryellow{\textbf{12.6}}                & \coloryellow{\textbf{15.5}}                 & \coloryellow{\textbf{16.8}}                         & 6.0                            & 8.0                        & 9.0                         \\
\textit{\textbf{3-shot}}                       & \coloryellow{\textbf{31.5}}                 & 34.1                          & 34.9                          & 9.3                          & 10.9                          & 11.5                         & \coloryellow{\textbf{9.6}}                   & \coloryellow{\textbf{11.7}}              & \coloryellow{\textbf{12.5}}                         \\ \hdashline 
\textit{\textbf{Relative Improvement}}         & {\color[HTML]{0500FF} -3.2\% $\downarrow$} & {\color[HTML]{FF0000} 13.5\% $\uparrow$} & {\color[HTML]{FF0000} 15.8\% $\uparrow$} & {\color[HTML]{0500FF} -16.7\% $\downarrow$} & {\color[HTML]{0500FF} -7.1\% $\downarrow$} & {\color[HTML]{0500FF} -2.4\% $\downarrow$} & {\color[HTML]{0500FF} -22.9\% $\downarrow$} & {\color[HTML]{0500FF} -5.1\% $\downarrow$} &  0\% \\ \bottomrule
\end{tabular}
\end{table*}

%% file: table/rq3tem_top.tex
\begin{table*}[]
\caption{Comparative evaluation of KareCoder with different ChatGPT Settings on various difficulties of CodeF post2021-9. For convenience of representation, we use ``Tem'' to represent ``Temperature''.}
\label{tab:6.3.2rq3tem_top_P}
\centering
\begin{tabular}{@{}lccccccccc@{}}
\toprule
                                                                                        & \multicolumn{3}{c}{\textit{\textbf{Simple}}}                                                 & \multicolumn{3}{c}{\textit{\textbf{Medium}}}                                                  & \multicolumn{3}{c}{\textit{\textbf{Hard}}}                                                     \\ \cmidrule(l){2-4} \cmidrule(l){5-7} \cmidrule(l){8-10}
\multirow{-2}{*}{\textit{\textbf{ChatGPT Setings}}}                                   & \textit{\textbf{Pass@1}}     & \textit{\textbf{Pass@3}}      & \textit{\textbf{Pass@5}}     & \textit{\textbf{Pass@1}}      & \textit{\textbf{Pass@3}}      & \textit{\textbf{Pass@5}}     & \textit{\textbf{Pass@1}}       & \textit{\textbf{Pass@3}}      & \textit{\textbf{Pass@5}}     \\ \midrule
\textit{\textbf{\begin{tabular}[c]{@{}l@{}}Tem=0.8 \hspace{0.15cm} Top\_P=0.95\end{tabular}}} & \coloryellow$29.4$                         & \coloryellow$33.7$                          & \coloryellow$35.4$                          & 8.7                           & \coloryellow$10.6$                          & \coloryellow$11.8$                          & \coloryellow{\textbf{8.6}}                   & \coloryellow{10.2}                          & \coloryellow{10.7}                          \\
\textit{\textbf{\begin{tabular}[c]{@{}l@{}}Tem=0.9 \hspace{0.15cm} Top\_P=0.95\end{tabular}}} & 29.3                & 31.5                 & 32.2                 & \coloryellow$8.8$                  & 10.5                 & 11.7                & 6.2                   & 7.6                  & 8.0                  \\
\textit{\textbf{\begin{tabular}[c]{@{}l@{}}Tem=1  \hspace{0.37cm} Top\_P=1\end{tabular}}}      & \colorblue{\textbf{30.5}}                & \colorblue{\textbf{38.8}}                 & \colorblue{\textbf{41.8}}                 & \colorblue{\textbf{10.5}}                 & \colorblue{\textbf{14.4}}                 & \colorblue{\textbf{16.4}}                 & \colorblue{7.4}                            & \colorblue{\textbf{11.1}}                 & \colorblue{\textbf{12.5}}                 \\ \hdashline 
\textit{\textbf{Relative Improvement}}                                                  & {\color[HTML]{FF0000} 3.7\% $\uparrow$} & {\color[HTML]{FF0000} 15.1\% $\uparrow$} & {\color[HTML]{FF0000} 18.1\% $\uparrow$} & {\color[HTML]{FF0000} 19.3\% $\uparrow$} & {\color[HTML]{FF0000} 35.8\% $\uparrow$} & {\color[HTML]{FF0000} 39.0\% $\uparrow$} & {\color[HTML]{0500FF} -14.0\% $\downarrow$} & {\color[HTML]{FF0000} 8.8\% $\uparrow$} & {\color[HTML]{FF0000} 16.8\% $\uparrow$} \\ \bottomrule
\end{tabular}
\end{table*}

%% file: table/rq4apps.tex
\begin{table}[]
\caption{Comparative evaluation of KareCoder with other methods on the top 500 problems in the APPS test set.}
\label{tab:6.4rq4apps}
\centering
\begin{tabular}{@{}lccc@{}}
\toprule
\textit{\textbf{Method}}               & \textit{\textbf{Pass@1}}      & \textit{\textbf{Pass@3}}     & \textit{\textbf{Pass@5}}     \\ \midrule
\textit{\textbf{ChatGPT}}              & 19.5                          & 30.0                         & 34.4                          \\ \midrule
\textit{\textbf{ChatGPT+Plan}}         & 9.2                           & 11.1                         & 11.9                          \\
\textit{\textbf{ChatGPT+SCOT}}          & \coloryellow\textbf{13.3}                          & 14.2                         & 14.4                          \\
\textit{\textbf{ChatGPT+SCOT\&KareCoder}}          & 13.0                          & \coloryellow$15.0$                          & \coloryellow\textbf{16.1}                          \\
\textit{\textbf{ChatGPT+KareCoder}}    & \colorblue$13.1$                 & \colorblue{\textbf{15.2}}                & \colorblue$15.8$                 \\ \hdashline 
\textit{\textbf{Relative Improvement}} & {\color[HTML]{0000FF} -1.5\% $\downarrow$} & {\color[HTML]{FF0000} 1.3\% $\uparrow$} & {\color[HTML]{0000FF} -1.9\% $\downarrow$} \\ \bottomrule
\end{tabular}
\end{table}

%% file: sec/7.threats_to_validity.tex
\section{Threats to Validity}\label{sec:7}

\subsection{Internal Validity}\label{sec:7.1}
\textbf{Limited algorithms and data structures tags.} Our experiments were conducted mainly on the CodeF dataset, the algorithms and data structures tags of CodeF are sourced from the CodeForces programming website. When testing on dataset with no tags, e.g., APPS, matching programing problems with the Knowledge Library is not feasible. To mitigate this issue, we devised prompts and 3-shot examples to aid ChatGPT in predicting tags, recommending suitable algorithms and data structures for problems lacking these tags. Although we have checked the tags generated by tag generator manually, the algorithms and data structures recommended may not always be the most fitting. The issue of recommending more appropriate algorithms and data structures for new problems remains an ongoing research challenge.

\textbf{Limited Knowledge Library and methods of integrating knowledge.} KareCoder has yet to achieve optimality in terms of the comprehensiveness of its Knowledge Library and the methods of knowledge integration. The current Knowledge Library equipped with KareCoder comprises information on algorithms and data structures. We use contextual prompts to integrate knowledge and problems and generate prompts that help solve programming problems. However, certain programming problems might necessitate using internet resources or extensions from Python's library. In the future, we intend to expand our Knowledge Library and enhance its quality, enabling KareCoder to access broader knowledge and constantly updating the ways in which KareCoder acquires knowledge to solve more complex programming problems.

\subsection{External Validity}\label{sec:7.2}
\textbf{Limited dataset.} At present, the specific training data of ChatGPT is not publicly available, leaving us uncertain if there is an overlap between the existing datasets and the training data of ChatGPT. If we use the training data as the test set for inference, the value of the research done would significantly diminish. Currently, we can only test using the CodeF dataset, and we are lacking other entirely new datasets to verify KareCoder's effectiveness in code generation. 
As the training data of ChatGPT gets updated, we may need to continually explore new data for research.

%% file: sec/7.disscussion.tex
\section{Disscuccion}
In this study, we assess the effectiveness of KareCoder by multiple RQs. Initially, we conducted preliminary exploration on ``gpt-3.5-turbo-0301'', which was followed by expanded experiments on ``gpt-3.5-turbo-0613''. However, our investigation was constrained by limited computational resources, preventing tests on a broader range of LLMs. We aim to extend our testing to additional LLMs to ascertain the scalability and practicality of KareCoder. The code generation dataset CodeF, proposed in this research, is based on the ChatGPT3.5 training data until September 2021. It is important to note that the existing dataset partitioning might lose relevance due to updates in the ChatGPT training data. To address this, we have included release timestamps for each problem in the dataset. Users are encouraged to download CodeF from our GitHub repository and can re-split the dataset to ensure it remains uncontaminated and relevant for their specific model requirements. Leveraging our Knowledge Library, researchers can apply the approach of integrating programming knowledge to diverse tasks. Nonetheless, it is essential to acknowledge that creating new datasets for other tasks, incorporating algorithms and data structures tags for these tasks, demands a certain amount of human resource investment.

%% file: sec/8.conclution.tex
\section{Conclusion}\label{sec:8}
In this research, we aimed to circumvent the utilization of LLMs' training data and thus constructed a new code generation dataset CodeF. The experiment indicated that problems predating September 2021 surpassed those following this date in terms of the Pass@k metric. We crafted a Knowledge Library specifically tailored for programming algorithm problems, and introduced KareCoder. This method integrates algorithms and data structures knowledge into the intermediate process of code generation, specifically, during the generation of knowledge-aware prompt. Experimental results demonstrated that KareCoder notably outperforms other code generation methods based on LLMs, on the CodeF post2021-9 and the first 500 problems of APPS datasets. This substantiates the research value of integrating algorithms and data structures knowledge into the code generation process.